%% file: root.tex
\documentclass[anon]{l4dc2026}

\usepackage[english]{babel}
\usepackage{array}
\usepackage{booktabs}   
\usepackage{amsfonts}
\newtheorem{assumption}{Assumption}
\newtheorem{problem}{Problem}

\title[Non-Myopic Safety Filter]{Non-Myopic Safety Filtering with Learned Intervention Cost-to-Go}
\usepackage{times}

\author{%
\Name{Johannes Autenrieb} \Email{johannes.autenrieb@dlr.de}\\
\Name{Mark Spiller} \Email{mark.spiller@dlr.de}\\
\addr German Aerospace Center (DLR)%
\AND
\Name{Daniel Silvestre} \Email{dsilvestre@fct.unl.pt}\\
\addr NOVA School of Science and Technology%
\AND
\Name{Namhoon Cho} \Email{namhoon.cho@snu.ac.kr}\\
\addr Seoul National University%
\AND
\Name{Jason J. Choi} \Email{jjhchoi@ucla.edu}\\
\addr University of California, Los Angeles%
}

\begin{document}

\maketitle

\thispagestyle{empty}

\begin{abstract}
\input{sections/Abstract}
\end{abstract}


\section{Introduction}
\input{./sections/Introduction}
\section{Preliminaries}
\input{sections/Prelimnimaries}

\section{Problem Formulation}
\label{sec:problem}
\input{sections/Problem}
\section{Non-Myopic Safety Filtering with Learned Intervention Cost-to-Go}
\label{sec:nonmyopic_safety_filter}
\input{sections/Main_Contribution}
\section{Simulation Results}
\label{sec:results}
\input{sections/Results}
\section{Conclusion}
\input{sections/Conclusion}

\bibliography{./references} 

\appendix
\input{sections/Appendix}

\end{document}

%% file: sections/Abstract.tex
Control barrier function (CBF) safety filters minimize instantaneous deviation from a nominal control command, but a sequence of pointwise-minimal interventions need not minimize cumulative intervention effort. This paper proposes a discrete-time non-myopic safety filter using a learned approximation of the optimal intervention cost-to-go over CBF-admissible policies. The approximation enters through a relaxed Bellman-residual constraint, while the original barrier condition remains hard. Feasibility is equivalent to that of the conventional CBF filter, and forward invariance remains independent of approximation accuracy under the stated assumptions. We derive a trajectory-wise performance certificate that requires neither the optimal value function nor an approximation-error bound. Under additional approximation and horizon assumptions, a suboptimality bound contains an explicit endpoint approximation-error term together with accumulated tolerance and relaxation terms. Numerical results for a collision-avoidance problem show an approximately $50\,\%$ reduction in mean cumulative intervention effort relative to conventional CBF filtering over $100$ initial conditions not used during training or parameter selection. The improvement is consistent across three independently trained networks, while maintaining the prescribed safety constraints and requiring less online computation than the tested predictive filter.

%% file: sections/Introduction.tex
Control barrier functions (CBFs) enforce safety constraints in nonlinear control systems by ensuring forward invariance of a prescribed safe set \citep{Ames_2019}. For control-affine systems, barrier conditions can often be incorporated into computationally efficient quadratic programs \citep{ames2014}. CBF-based safety filters minimally modify a controller command to satisfy safety constraints, with applications in robotics, autonomous systems, and aerospace control \citep{Shi2024,autenrieb2025_JGCD,Hafner2026,autenrieb2025_ACC}. Discrete-time formulations extend this approach to discrete-time dynamics \citep{Agrawal2017,Zeng2021}.

Conventional CBF safety filters minimize the instantaneous deviation from the nominal input without considering the interventions required at subsequent steps. Consequently, applying a currently admissible command can lead to states requiring substantially larger or repeated corrections later. This motivates accounting for future intervention costs when selecting the current control input.

Predictive safety filters use online trajectory optimization to anticipate future constraint violations \citep{wabersich2021predictive}. Predictive CBFs characterize constraint recoverability through value functions \citep{Wabersich2023}, and learned approximations can reduce the need for online optimization \citep{Didier2026}. Predictive stability filters additionally impose Lyapunov decrease conditions \citep{Didier2024}.

\citet{GrosZanonBemporad2020} select safe actions by minimizing a learned action-value function over the admissible input set. \citet{LinXu2026} use action-value curvature to weight the projection of a reference input onto this set, leaving an already admissible reference unchanged. \citet{ShahrakiLessard2026} minimize a pre-Hamiltonian derived from an unconstrained task-value approximation under hard barrier constraints. In contrast, our value function represents the minimum cumulative deviation from a fixed nominal controller over barrier-admissible policies. It therefore accounts for the intervention costs induced by the safety constraints and allows the filter to consider preventive corrections even when the current command is admissible.

We incorporate the learned intervention cost-to-go through a relaxed Bellman-residual condition in a discrete-time CBF filter. The original barrier constraints remain hard, while an unbounded nonnegative slack preserves feasibility whenever the conventional CBF filter is feasible. Under the stated assumptions, forward invariance is therefore independent of the approximation accuracy. We establish a trajectory-wise bound on cumulative intervention costs and, under additional approximation and horizon assumptions, a suboptimality bound relative to the barrier-admissible optimum. The latter contains an endpoint approximation-error term together with accumulated tolerance and relaxation terms. In a fixed collision-avoidance environment, the proposed filter reduces mean cumulative intervention cost by approximately $50\,\%$ relative to conventional CBF filtering across $100$ initial conditions not used during training or parameter selection. The improvement is consistent across three independently trained networks.

%% file: sections/Prelimnimaries.tex
\subsection{Control Barrier Functions}
\label{subsec:cbf_preliminaries}

Consider the discrete-time nonlinear control-affine system
\begin{equation}
\label{eq:cbf_system}
x_{k+1}=f(x_k)+g_u(x_k)u_k,
\end{equation}
where $x_k\in\mathcal X\subseteq\mathbb R^n$ and $u_k\in U\subseteq\mathbb R^m$ denote the state and control input, respectively. The mappings $f:\mathcal X\rightarrow\mathbb R^n$ and $g_u:\mathcal X\rightarrow\mathbb R^{n\times m}$ are assumed to be locally Lipschitz continuous.

Safety is characterized by a continuously differentiable function $h:\mathcal X\rightarrow\mathbb R$, defining the safe set
\begin{equation}
\label{eq:safe_set}
\mathcal C=\left\{x\in\mathcal X:h(x)\ge0\right\}.
\end{equation}
The set $\mathcal C$ is forward invariant if every trajectory initialized in $\mathcal C$ remains in $\mathcal C$ for all subsequent time steps (\cite{Blanchini_1999}). A discrete-time control barrier function (CBF) provides a sufficient condition for forward invariance by requiring that, for every $x_k\in\mathcal C$, there exists an admissible control input satisfying (\cite{Xiong2023})
\begin{equation}
\label{eq:cbf_condition}
h(x_{k+1})\ge(1-\gamma)h(x_k),
\qquad \gamma\in(0,1].
\end{equation}
The corresponding set of safety-preserving control inputs is
\begin{equation}
\label{eq:safe_input_set}
K_{\mathcal C}(x_k)=
\left\{
u_k\in U:
h\big(f(x_k)+g_u(x_k)u_k\big)
\ge(1-\gamma)h(x_k)
\right\}.
\end{equation}
If $K_{\mathcal C}(x)\neq\emptyset$ for all $x\in\mathcal C$, any controller satisfying $u_k\in K_{\mathcal C}(x_k)$ renders $\mathcal C$ forward invariant by recursive application of \eqref{eq:cbf_condition}.

Given a nominal control command $u_{\mathrm{nom},k}$, the conventional CBF safety filter selects the admissible input minimizing the instantaneous intervention effort:
\begin{equation}
\label{eq:cbf_qp}
\begin{aligned}
u_k^{\mathrm{CBF}}
\in\arg\min_{u_k\in U}\quad&
\frac{1}{2}\left\|u_k-u_{\mathrm{nom},k}\right\|_2^2\\
\textnormal{s.t.}\quad&
u_k\in K_{\mathcal C}(x_k).
\end{aligned}
\end{equation}
The optimization problem \eqref{eq:cbf_qp} is pointwise optimal with respect to intervention effort. However, it does not explicitly account for how the current control action influences the interventions required at subsequent time steps.

\subsection{Cumulative Intervention Cost}
\label{subsec:intervention_cost}

Consider a nominal state-feedback controller $u_{\mathrm{nom},k}=\kappa(x_k)$ and define the intervention stage cost as
\begin{equation}
\label{eq:intervention_cost}
\ell(x_k,u_k)=
\left\|u_k-\kappa(x_k)\right\|_2^2.
\end{equation}
Let $\pi=(\pi_0,\ldots,\pi_{N-1})$ be a feedback policy over a horizon $N$, with $\pi_k:\mathcal X\to U$. Starting from $x_0^\pi=x_0$, the policy selects $u_k^\pi=\pi_k(x_k^\pi)$, and \eqref{eq:cbf_system} determines $x_{k+1}^\pi$ from $(x_k^\pi,u_k^\pi)$. The policy is barrier-admissible from $x_0$ if $u_k^\pi\in K_{\mathcal C}(x_k^\pi)$ for every $k=0,\ldots,N-1$. Its cumulative intervention cost is
\begin{equation}
\label{eq:performance_cost}
J_N(\pi;x_0)=
\sum_{k=0}^{N-1}\ell(x_k^\pi,u_k^\pi).
\end{equation}

The optimal cumulative intervention cost is
\begin{equation}
\label{eq:optimal_intervention_cost}
J_N^\star(x_0)=
\inf_{\pi} J_N(\pi;x_0),
\end{equation}
where the infimum is taken over policies that are barrier-admissible from $x_0$. This benchmark accounts for how each input affects future intervention costs. Consequently, the pointwise CBF filter \eqref{eq:cbf_qp} need not attain $J_N^\star(x_0)$, motivating the proposed non-myopic formulation.

%% file: sections/Problem.tex
Let $\pi_{\mathrm{CBF}}$ denote the policy induced by the conventional CBF filter \eqref{eq:cbf_qp}. Its pointwise minimization of intervention effort can yield a cumulative cost $J_N(\pi_{\mathrm{CBF}};x_0)$ above the barrier-admissible benchmark $J_N^\star(x_0)$ in \eqref{eq:optimal_intervention_cost}. The benchmark is restricted to policies satisfying the prescribed barrier condition.

A nominal input can be barrier-admissible at the current state and still lead to states that require substantial future interventions. We therefore seek to approximate the optimal intervention cost-to-go for a fixed nominal controller under the prescribed barrier conditions. These conditions enter the definition of the learning target, whereas \citet{ShahrakiLessard2026} use an unconstrained task-value function.

\begin{problem}[Non-myopic safety filtering]
\label{prob:nonmyopic_safety_filtering}
Given the discrete-time dynamics $x_{k+1}=f(x_k)+g_u(x_k)u_k$ in \eqref{eq:cbf_system}, the safe set $\mathcal C$, and a fixed nominal controller $u_{\mathrm{nom},k}=\kappa(x_k)$, design a safety filter that reduces the cumulative intervention cost \eqref{eq:performance_cost}, relative to $\pi_{\mathrm{CBF}}$ governed by \eqref{eq:cbf_qp}. The filter must enforce $u_k\in K_{\mathcal C}(x_k)$ as a hard constraint at each step and avoid online optimization over a multi-step input sequence. Under the stated model and feasibility assumptions, it must preserve forward invariance of $\mathcal C$ independently of the accuracy of the learned cost-to-go.
\end{problem}

%% file: sections/Main_Contribution.tex
The proposed filter incorporates learned future intervention costs while retaining the original CBF safety constraint. The corresponding feasibility, safety, and performance guarantees are established below, with proofs provided in Appendix~\ref{app:nonmyopic_proofs}.

\subsection{Learned Future-Cost Approximation}
\label{subsec:learned_future_cost}

Consider the time-invariant nominal state-feedback law $u_{\mathrm{nom},k}=\kappa(x_k)$ introduced in Section~\ref{subsec:intervention_cost}, such that the intervention cost \eqref{eq:intervention_cost} depends only on the current state and control input. For $x\in\mathcal C$, define the safety-constrained optimal cost-to-go
\begin{equation}
\label{eq:safety_cost_to_go}
V^\star(x)
=
\inf_{\{u_i\}_{i=0}^{\infty}}
\sum_{i=0}^{\infty}
\ell(x_i,u_i),
\qquad x_0=x,
\end{equation}
subject to \eqref{eq:cbf_system} and $u_i\in K_{\mathcal C}(x_i)$ for all $i\ge0$. The function $V^\star$ represents the minimum cumulative intervention effort achievable from a given state while satisfying the prescribed CBF condition. Since the stage cost is nonnegative, $V^\star(x)\ge0$. In the following, the optimal cost is assumed finite on the states of interest, and the infimum in \eqref{eq:safety_cost_to_go} is assumed to be attained whenever an optimal action is invoked.

\begin{assumption}[Horizon completeness]
\label{ass:terminal_inactivity}
The evaluation horizon $N$ is sufficiently long that the finite-horizon optimal intervention cost coincides with the infinite-horizon optimum, i.e.,
\begin{equation}
\label{eq:jstar_vstar}
J_N^\star(x_0)=V^\star(x_0).
\end{equation}
A sufficient condition for \eqref{eq:jstar_vstar} is that a finite-horizon optimal input sequence admits an infinite continuation satisfying the prescribed CBF condition with zero intervention cost for all $i\ge N$.
\end{assumption}

Assumption~\ref{ass:terminal_inactivity} relates the finite-horizon benchmark \eqref{eq:optimal_intervention_cost} to the infinite-horizon value function. By the principle of optimality, $V^\star$ satisfies
\begin{equation}
\label{eq:bellman_inequality}
V^\star(x_{k+1})
-
V^\star(x_k)
+
\ell(x_k,u_k)
\ge0
\end{equation}
for every $u_k\in K_{\mathcal C}(x_k)$. Equality holds for a Bellman-optimal action $u_k^{\mathrm{opt}}$. In the residuals below, $x_{k+1}$ denotes the successor predicted by \eqref{eq:cbf_system} for the input being evaluated.

Direct evaluation of $V^\star$ generally requires solving a constrained optimal control problem. The proposed method therefore introduces a learned approximation
\begin{equation}
\label{eq:learned_future_cost}
W_\theta:\mathcal C\rightarrow\mathbb R_{\ge0},
\end{equation}
where $\theta$ denotes the parameters of a function approximator, such as a neural network, obtained offline from safety-filtering data, expert demonstrations, or optimization-based methods, in line with related approaches \citep{Didier2026,Wabersich2023,Lindemann2024,Zeng2021}. Nonnegativity reflects the interpretation of $W_\theta$ as an estimate of future intervention cost. Once trained, the function can be evaluated online without solving a predictive optimal control problem.

\begin{definition}[Learned future-cost approximation]
\label{def:learned_future_cost}
A continuously differentiable function $W_\theta:\mathcal C\rightarrow\mathbb R_{\ge0}$ is called a learned future-cost approximation if it approximates the safety-constrained optimal cost-to-go $V^\star$ defined in \eqref{eq:safety_cost_to_go}.
\end{definition}

For the benchmark-relative performance analysis, we assume a uniform approximation error.

\begin{assumption}[Uniform approximation error]
\label{ass:uniform_approximation}
There exist a compact set $\mathcal X_r\subseteq\mathcal C$ containing the trajectories considered in the performance analysis and a constant $\varepsilon_W\ge0$ such that
\begin{equation}
\label{eq:uniform_approximation}
\left|W_\theta(x)-V^\star(x)\right|
\le\varepsilon_W,
\qquad \forall x\in\mathcal X_r.
\end{equation}
\end{assumption}

To relate the learned approximation to the optimal control problem, define the Bellman residual
\begin{equation}
\label{eq:bellman_residual}
r^\star(x_k,u_k)
=
V^\star(x_{k+1})
-
V^\star(x_k)
+
\ell(x_k,u_k).
\end{equation}
By \eqref{eq:bellman_inequality}, $r^\star(x_k,u_k)\ge0$ for every admissible action, with equality for a Bellman-optimal action. The residual therefore quantifies the additional intervention cost associated with the current action relative to an optimal continuation.

Replacing $V^\star$ by $W_\theta$ yields the approximate Bellman residual
\begin{equation}
\label{eq:learned_bellman_residual}
r_\theta(x_k,u_k)
=
W_\theta(x_{k+1})
-
W_\theta(x_k)
+
\ell(x_k,u_k).
\end{equation}
Under Assumption~\ref{ass:uniform_approximation}, the residual approximation error satisfies
\begin{equation}
\label{eq:bellman_residual_error}
\left|
r_\theta(x_k,u_k)
-
r^\star(x_k,u_k)
\right|
\le2\varepsilon_W
\end{equation}
whenever $x_k,x_{k+1}\in\mathcal X_r$. In particular, if both $x_k$ and the successor under a Bellman-optimal action $u_k^{\mathrm{opt}}$ belong to $\mathcal X_r$, then
\begin{equation}
\label{eq:optimal_learned_residual}
r_\theta(x_k,u_k^{\mathrm{opt}})
\le2\varepsilon_W.
\end{equation}

These relations motivate selecting control actions whose approximate Bellman residual remains below a prescribed tolerance $\eta\ge0$, namely
\begin{equation}
\label{eq:learned_residual_condition}
W_\theta(x_{k+1})
-
W_\theta(x_k)
+
\ell(x_k,u_k)
\le\eta.
\end{equation}
When $W_\theta=V^\star$, setting $\eta=0$ restricts \eqref{eq:learned_residual_condition} to Bellman-optimal actions. Under Assumption~\ref{ass:uniform_approximation}, if $\eta\ge2\varepsilon_W$ and both $x_k$ and the successor under a Bellman-optimal input belong to $\mathcal X_r$, that input satisfies both the barrier condition and \eqref{eq:learned_residual_condition}. Thus, a feasible input with zero slack exists at $x_k$ under these conditions.

\subsection{Proposed Non-Myopic Safety Filter}
\label{subsec:proposed_filter}

When the tolerance is too small relative to the learned residuals, \eqref{eq:learned_residual_condition} can exclude every input in a nonempty $K_{\mathcal C}(x_k)$. To preserve feasibility whenever the conventional CBF filter is feasible, we relax the residual condition and enforce barrier admissibility as a hard constraint. At each time step $k$, the filter solves
\begin{equation}
\label{eq:nonmyopic_safety_filter}
\begin{aligned}
(u_k^\star,\delta_k^\star)
\in\arg\min_{u,\delta}\quad&
\frac{1}{2}\|u-u_{\mathrm{nom},k}\|_2^2
+\frac{\lambda}{2}\delta^2\\
\textnormal{s.t.}\quad&
u\in K_{\mathcal C}(x_k),\\
& r_\theta(x_k,u)\le\eta+\delta,\qquad \delta\ge0.
\end{aligned}
\end{equation}
Here, $x_k$ and $u_{\mathrm{nom},k}=\kappa(x_k)$ are fixed, and the candidate successor used in $r_\theta(x_k,u)$ is $f(x_k)+g_u(x_k)u$. The filter applies $u_k^\star$, and $\delta_k^\star$ is the realized slack. The decision variable $\delta$ has no upper bound, and $\lambda>0$ weights residual exceedance against immediate intervention. For finite $\lambda$, the optimizer may select positive slack even when a zero-slack input exists. For a fixed CBF-admissible input $u$, the optimal slack is
\begin{equation}
\label{eq:optimal_slack_given_input}
\delta_k^{\mathrm{opt}}(u)=\max\{0,r_\theta(x_k,u)-\eta\}.
\end{equation}
Substituting \eqref{eq:optimal_slack_given_input} into \eqref{eq:nonmyopic_safety_filter} gives an equivalent input minimization:
\begin{equation}
\label{eq:nonmyopic_filter_reduced}
\begin{aligned}
u_k^\star\in\arg\min_{u\in K_{\mathcal C}(x_k)}\quad&
\frac{1}{2}\|u-u_{\mathrm{nom},k}\|_2^2
+\frac{\lambda}{2}[\delta_k^{\mathrm{opt}}(u)]^2.
\end{aligned}
\end{equation}
The penalty is zero for residuals at or below $\eta$. Direct learned-value lookahead ranks admissible inputs by the sum of immediate and predicted future intervention costs:
\begin{equation}
\label{eq:direct_learned_value_filter}
u_k^{\mathrm{LA}}\in\arg\min_{u\in K_{\mathcal C}(x_k)}
\bigl\{\ell(x_k,u)+W_\theta(f(x_k)+g_u(x_k)u)\bigr\}.
\end{equation}
The two formulations use the learned cost-to-go differently. While \eqref{eq:direct_learned_value_filter} directly minimizes the sum of current and predicted future intervention costs, \eqref{eq:nonmyopic_filter_reduced} retains the instantaneous intervention objective and penalizes Bellman-residual violations beyond the tolerance $\eta$. If the conventional CBF input satisfies the residual condition, the proposed filter recovers that input, whereas direct lookahead may still select a different action. The relaxed residual condition also provides the basis for the trajectory-wise performance certificate in Lemma~\ref{lem:certificate}. Appendix~\ref{app:lookahead_comparison} compares the resulting performance bounds.

If $u_{\mathrm{nom},k}\in K_{\mathcal C}(x_k)$, the conventional myopic safety filter in \eqref{eq:cbf_qp} returns the nominal command with zero immediate intervention cost. If the nominal command's learned residual exceeds $\eta$, our proposed filter may instead intervene \textit{preventively}, accepting a larger immediate correction to reduce the cumulative intervention cost over time. Further implementation details are given in Appendix~\ref{app:nonmyopic_remarks}.

\subsection{Feasibility and Safety Preservation}
\label{subsec:safety_preservation}

\begin{lemma}[Feasibility equivalence]
\label{lem:feasibility}
For every $x_k\in\mathcal C$, problem \eqref{eq:nonmyopic_safety_filter} is feasible if and only if \eqref{eq:cbf_qp} is feasible, i.e., if and only if $K_{\mathcal C}(x_k)\neq\emptyset$.
\end{lemma}

The proof is provided in Appendix~\ref{app:proof_feasibility}.

\begin{theorem}[Safety preservation]
\label{thm:learned_filter_safety}
Suppose $K_{\mathcal C}(x)$ is nonempty for all $x\in\mathcal C$, $h$ satisfies \eqref{eq:cbf_condition}, and the optimization problem \eqref{eq:nonmyopic_safety_filter} admits a minimizer at each state in $\mathcal C$. Then the resulting control law renders $\mathcal C$ forward invariant, independently of the approximation quality of $W_\theta$.
\end{theorem}

The proof is provided in Appendix~\ref{app:proof_safety}.

\subsection{Cumulative Intervention Performance}
\label{subsec:nonmyopic_performance}

We first bound the cumulative intervention cost along a realized trajectory without requiring an approximation-error bound. The relation to the barrier-admissible optimum is then established under Assumptions~\ref{ass:terminal_inactivity} and~\ref{ass:uniform_approximation}.

\begin{lemma}[Pathwise performance certificate]
\label{lem:certificate}
Let $\{x_k\}_{k=0}^{N}$ be a trajectory generated by \eqref{eq:nonmyopic_safety_filter}, and define
\begin{equation}
\label{eq:j_nmf}
J_{\mathrm{NMF}}
=
\sum_{k=0}^{N-1}
\ell(x_k,u_k^\star).
\end{equation}
Then
\begin{equation}
\label{eq:pathwise_certificate}
J_{\mathrm{NMF}}
\le
W_\theta(x_0)
-
W_\theta(x_N)
+
N\eta
+
\sum_{k=0}^{N-1}\delta_k^\star.
\end{equation}
\end{lemma}

The proof is provided in Appendix~\ref{app:proof_certificate}. If $r_\theta(x_k,u_k^\star)\le\eta$ at every step, then $\delta_k^\star=0$ throughout the trajectory by \eqref{eq:optimal_slack_given_input}, and the slack sum vanishes from the performance bounds.

\begin{theorem}[Suboptimality bound]
\label{thm:nonmyopic_performance_bound}
Suppose Assumption~\ref{ass:uniform_approximation} holds along the trajectory of the proposed filter. Then
\begin{equation}
\label{eq:approx_nonmyopic_bound}
\begin{aligned}
J_{\mathrm{NMF}}
\le{}&
V^\star(x_0)-V^\star(x_N)
+
2\varepsilon_W
+
N\eta + \sum_{k=0}^{N-1}\delta_k^\star.
\end{aligned}
\end{equation}
If Assumption~\ref{ass:terminal_inactivity} also holds, then
\begin{equation}
\label{eq:approx_nonmyopic_bound_terminal}
J_{\mathrm{NMF}}
\le
J_N^\star(x_0)
+
2\varepsilon_W
+
N\eta
+
\sum_{k=0}^{N-1}\delta_k^\star.
\end{equation}
\end{theorem}

The proof is provided in Appendix~\ref{app:proof_suboptimality}. The explicit value-approximation term is bounded by the endpoint contribution $2\varepsilon_W$. However, the tolerance and accumulated slack may themselves depend on approximation quality and grow with the horizon; the bound therefore does not establish horizon-independent sensitivity of cumulative intervention cost to approximation error.

\begin{corollary}[Suboptimality ratio bound]
\label{cor:suboptimality_ratio_bound}
Under Assumptions~\ref{ass:terminal_inactivity} and~\ref{ass:uniform_approximation}, with $J_N^\star(x_0)>0$, the realized suboptimality ratio satisfies
\begin{equation}
\label{eq:suboptimality_ratio_bound}
\rho_{\mathrm{NMF}}
=
\frac{J_{\mathrm{NMF}}}{J_N^\star(x_0)}
\le
1+
\frac{
2\varepsilon_W
+
N\eta
+
\sum_{k=0}^{N-1}\delta_k^\star
}
{J_N^\star(x_0)}.
\end{equation}
\end{corollary}

The bound in Corollary~\ref{cor:suboptimality_ratio_bound} depends on the realized cumulative slack and therefore does not establish a uniform suboptimality ratio without an additional bound on the relaxation relative to $J_N^\star(x_0)$.

\subsection{Comparison with the Conventional CBF Filter}
\label{subsec:performance_improvement}

Let $\pi_{\mathrm{CBF}}$ and $\pi_{\mathrm{NMF}}$ denote the policies induced by \eqref{eq:cbf_qp} and \eqref{eq:nonmyopic_safety_filter}, respectively. Throughout this comparison, both policies use the same initial condition $x_0$ and evaluation horizon $N$, and we write $J(\pi)=J_N(\pi;x_0)$. To compare their cumulative intervention costs, we express the cost of conventional CBF filtering in terms of the learned Bellman residual. For a policy $\pi$, define the accumulated residual
\begin{equation}
\label{eq:learned_residual}
R_\theta(\pi)
=
\sum_{k=0}^{N-1}
\left[
W_\theta(x_{k+1}^{\pi})
-
W_\theta(x_k^{\pi})
+
\ell(x_k^{\pi},u_k^{\pi})
\right].
\end{equation}

\begin{lemma}[Residual identity]
\label{lem:residual_identity}
For every policy $\pi$ and every realized trajectory,
\begin{equation}
\label{eq:residual_identity}
R_\theta(\pi)
=
W_\theta(x_N^{\pi})
-
W_\theta(x_0)
+
J(\pi).
\end{equation}
\end{lemma}

The proof is provided in Appendix~\ref{app:proof_residual_identity}. The identity expresses the cumulative intervention cost of conventional CBF filtering through its accumulated learned residual and terminal value. Combining this relation with the upper bound on NMF intervention cost in Lemma~\ref{lem:certificate} yields the sufficient improvement condition in Theorem~\ref{thm:performance_improvement}.

\begin{theorem}[Comparison with the conventional filter]
\label{thm:performance_improvement}
Let both policies start from the same initial condition and define the terminal discrepancy
\begin{equation}
\label{eq:terminal_value_comparison}
\Delta_W
=
W_\theta(x_N^{\mathrm{CBF}})
-
W_\theta(x_N^{\mathrm{NMF}}).
\end{equation}
Then
\begin{equation}
\label{eq:performance_improvement_bound}
\begin{aligned}
&J(\pi_{\mathrm{NMF}})
-
J(\pi_{\mathrm{CBF}})\le
\Delta_W
-
\left[
R_\theta(\pi_{\mathrm{CBF}})
-
N\eta
-
\sum_{k=0}^{N-1}\delta_k^\star
\right].
\end{aligned}
\end{equation}
Consequently, the proposed filter attains a strictly smaller cumulative intervention cost than $\pi_{\mathrm{CBF}}$ whenever
\begin{equation}
\label{eq:residual_improvement_condition}
R_\theta(\pi_{\mathrm{CBF}})
>
N\eta
+
\sum_{k=0}^{N-1}\delta_k^\star
+
\Delta_W.
\end{equation}
\end{theorem}

The proof is provided in Appendix~\ref{app:proof_improvement}. Using Lemma~\ref{lem:residual_identity}, condition~\eqref{eq:residual_improvement_condition} can equivalently be written as
\[
J(\pi_{\mathrm{CBF}})>W_\theta(x_0)-W_\theta(x_N^{\mathrm{NMF}})+N\eta+\sum_{k=0}^{N-1}\delta_k^\star.
\]
The right-hand side provides an upper bound on the realized NMF intervention cost by Lemma~\ref{lem:certificate}. Thus, improvement is certified when the cumulative intervention cost of conventional CBF filtering exceeds this bound. This can occur when pointwise-minimal corrections lead to larger or repeated interventions later in the trajectory. The condition is sufficient and can be checked after both trajectories have been obtained.

%% file: sections/Results.tex
We evaluate the proposed non-myopic safety filter (NMF) in a three-dimensional obstacle-avoidance task using a simplified nonlinear model of quadrotor translational dynamics. The model assumes ideal thrust-vector tracking and includes quadratic drag. For $x_k=(p_k,v_k)$ and gravity-compensated acceleration command $u_k$, the sampled dynamics are
\begin{equation}
\label{eq:quadrotor_dynamics}
\begin{aligned}
p_{k+1}&=p_k+\Delta t\,v_k,\\
v_{k+1}&=v_k+\Delta t\bigl(u_k-c_d\|v_k\|_2v_k\bigr),
\end{aligned}
\end{equation}
where $\Delta t=0.15\,\mathrm{s}$ and $c_d=0.12\,\mathrm{m}^{-1}$. The nominal controller is a saturated PD law, $\kappa(x)=\operatorname{proj}_U[-K_p(p-p_{\mathrm g})-K_dv]$, with $K_p=0.55\,\mathrm{s}^{-2}I_{3\times 3}$ and $K_d=1.35\,\mathrm{s}^{-1}I_{3\times 3}$. Two spherical obstacles define barriers based on clearance and radial velocity. Since the Euler update fixes $p_{k+1}$ at a given state, the sampled barrier inequalities are affine in $u_k$ (Appendix~\ref{app:results_setup}).

All controllers use the same model, input bounds, and hard barrier inequalities. The conventional CBF filter minimizes $\ell(x_k,u_k)=\|u_k-\kappa(x_k)\|_2^2$. The H-CBF-QP of \citet{ShahrakiLessard2026} projects a Hamiltonian-based command onto the same admissible set. MPC-CBF minimizes the sum of stage costs over a $15$-step prediction. We evaluate cumulative intervention as $J_N=\sum_{k=0}^{N-1}\ell(x_k,u_k)$ over $N=90$ steps. The primary NMF network (seed $11$) is fitted to $7{,}619$ continuation-cost labels from $800$ training starts. The parameters $(\lambda,\eta)=(20,0)$ are selected on $40$ separate validation starts. Performance is evaluated on $100$ initial conditions not used during training or parameter selection. Three networks are trained independently with different random initializations to assess the consistency of the results. All splits share the obstacle geometry. Appendix~\ref{app:results_training} describes the label generation and the numerical offline optimal solution.

\begin{figure}[t]
\centering
\includegraphics[width=0.70\columnwidth]{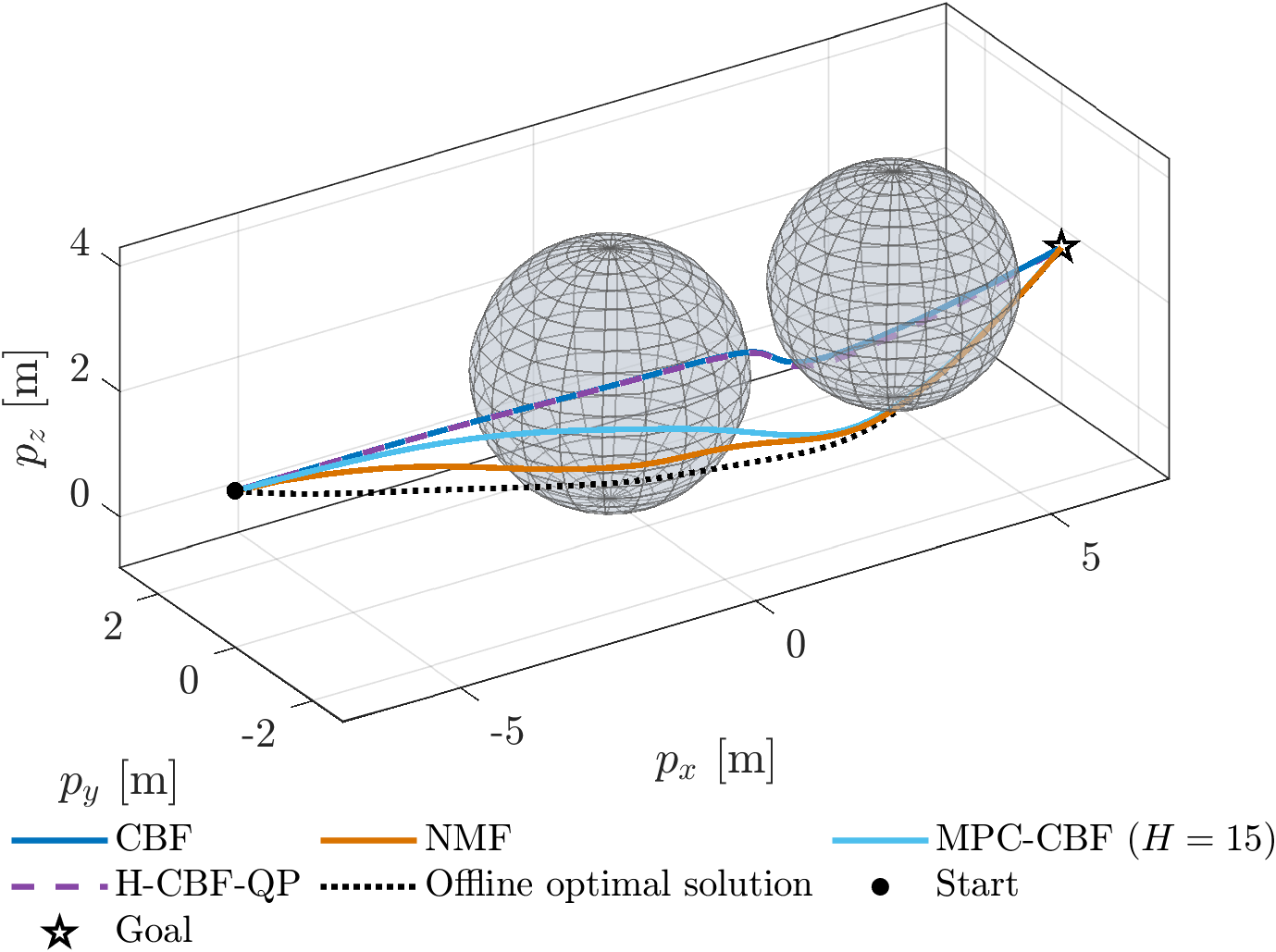}
\caption{Controller trajectories and offline optimal solution.}
\label{fig:trajectory}
\end{figure}

Figure~\ref{fig:trajectory} shows a representative case starting from rest at $p_0=(-7,0,1.7)^\top\,\mathrm m$.

At the initial state in Fig.~\ref{fig:trajectory}, the nominal command is barrier-admissible and CBF applies it with zero stage cost, yielding a learned residual of $2.04$. NMF changes the successor velocity at a stage cost of $0.72$ and reduces the learned cost-to-go by $0.70$ relative to the initial state. The resulting residual is $0.014$. For this start, NMF reduces $J_N$ by $57.1\,\%$ relative to CBF and incurs $9.4\,\%$ higher cost than MPC-CBF (Table~\ref{tab:quadrotor_results}). The minimum sampled clearance increases from $0.062\,\mathrm m$ for CBF to $0.217\,\mathrm m$ for NMF. Appendix~\ref{app:results_trajectories} reports the clearance and intervention histories. The a posteriori bound $J_{\mathrm{NMF}}\le100.67$ certifies improvement over CBF for this start (Appendix~\ref{app:results_certificate}).

\begin{table}[t]
\centering
\caption{Intervention costs, savings relative to CBF, and mean online time over $100$ test cases.}
\label{tab:quadrotor_results}
\small
\setlength{\tabcolsep}{4pt}
\begin{tabular}{lrrrr}
\hline
Method & $J_N$ (Fig.~\ref{fig:trajectory}) & Mean $J_N$ &
Saving vs.\ CBF & Time [ms]\\
\hline
CBF & 218.0 & 200.5 & -- & 0.31\\
H-CBF-QP & 209.8 & 188.8 & 11.7 (5.8\%) & 0.50\\
MPC-CBF ($H=15$) & 85.5 & 83.8 &
116.7 (58.2\%) & 24.97\\
\textbf{NMF} & \textbf{93.6} & \textbf{101.3} & \textbf{99.2 (49.5\%)} & \textbf{3.31}\\
Offline optimal solution & 74.4 & -- & -- & --\\
\hline
\end{tabular}
\end{table}

Across the $100$ test cases, NMF reduces mean cumulative intervention cost by $49.5\,\%$ relative to CBF and $46.4\,\%$ relative to H-CBF-QP. NMF achieves a lower cost than both filters for every tested initial condition. All controllers complete the evaluation horizon safely and reach the goal. The corresponding reductions relative to CBF for the other two independently trained networks are $50.6\,\%$ and $49.2\,\%$, respectively, showing consistent performance across different network initializations.

The mean online time per step for NMF is about $13\,\%$ of that for MPC-CBF over the test cases. With the slack eliminated, NMF optimizes over one three-dimensional command, while MPC-CBF optimizes over $45$ input components. Appendix~\ref{app:results_timing} reports the distributions of cumulative intervention cost and online computation time, together with solver details and observed maximum times.

%% file: sections/Conclusion.tex
This paper proposed a non-myopic CBF safety filter that uses a learned intervention cost-to-go through a relaxed Bellman-residual constraint. The filter retains the original barrier-admissibility conditions, preserving feasibility and safety independently of approximation accuracy under the stated assumptions. The performance analysis provides a trajectory-wise certificate and a conditional suboptimality bound with an explicit endpoint approximation-error term. Accumulated tolerance and relaxation remain in the bound, so it does not guarantee improvement over conventional filtering. In the evaluated collision-avoidance problem, mean intervention effort decreased by approximately $50\,\%$ relative to conventional CBF filtering across $100$ test cases, with consistent results from three independently trained networks and less online computation than the tested predictive baseline. Future work will examine broader safety environments and adaptation under model uncertainty.

%% file: sections/Appendix.tex
\section{Proofs and Supplementary Discussion}
\label{app:nonmyopic_proofs}

\subsection{Proof of Feasibility Equivalence}
\label{app:proof_feasibility}
\begin{proof}
Necessity follows because every feasible input of \eqref{eq:nonmyopic_safety_filter} belongs to $K_{\mathcal C}(x_k)$. For sufficiency, let $u\in K_{\mathcal C}(x_k)$ and choose
\begin{equation}
\delta
=
\max\left\{
0,\,
W_\theta(f(x_k)+g_u(x_k)u)
-
W_\theta(x_k)
+
\ell(x_k,u)
-
\eta
\right\}.
\end{equation}
Since $W_\theta$ is finite-valued and $\delta$ is unbounded above, $(u,\delta)$ satisfies both constraints of \eqref{eq:nonmyopic_safety_filter}.
\end{proof}

\subsection{Proof of Safety Preservation}
\label{app:proof_safety}
\begin{proof}
By the assumed feasibility of the discrete-time CBF condition, $K_{\mathcal C}(x_k)\neq\emptyset$ for every $x_k\in\mathcal C$. Lemma~\ref{lem:feasibility} therefore establishes feasibility of \eqref{eq:nonmyopic_safety_filter}. Every optimizer satisfies $u_k^\star\in K_{\mathcal C}(x_k)$, and forward invariance follows by induction from $h(x_{k+1})\ge(1-\gamma)h(x_k)\ge0$ whenever $x_k\in\mathcal C$. The learned constraint is unconditionally relaxable and cannot invalidate the hard barrier condition.
\end{proof}

\subsection{Proof of Pathwise Performance Certificate}
\label{app:proof_certificate}
\begin{proof}
Evaluating the learned constraint in \eqref{eq:nonmyopic_safety_filter} at the optimizer gives
\begin{equation}
\label{eq:learned_constraint_optimizer}
W_\theta(x_{k+1})
-
W_\theta(x_k)
+
\ell(x_k,u_k^\star)
\le
\eta+\delta_k^\star.
\end{equation}
Summing over $k=0,\ldots,N-1$, the learned-value terms telescope, yielding $W_\theta(x_N)-W_\theta(x_0)+J_{\mathrm{NMF}}\le N\eta+\sum_{k=0}^{N-1}\delta_k^\star$, which rearranges to \eqref{eq:pathwise_certificate}.
\end{proof}

\subsection{Proof of Suboptimality Bound}
\label{app:proof_suboptimality}
\begin{proof}
Applying \eqref{eq:uniform_approximation} to the endpoint terms of \eqref{eq:pathwise_certificate} gives $W_\theta(x_0)\le V^\star(x_0)+\varepsilon_W$ and $-W_\theta(x_N)\le -V^\star(x_N)+\varepsilon_W$, establishing \eqref{eq:approx_nonmyopic_bound}. Under Assumption~\ref{ass:terminal_inactivity}, $V^\star(x_0)=J_N^\star(x_0)$. Since $V^\star(x_N)\ge0$, dropping the nonpositive terminal term yields \eqref{eq:approx_nonmyopic_bound_terminal}.
\end{proof}

\subsection{Proof of Residual Identity}
\label{app:proof_residual_identity}
\begin{proof}
The learned-value terms in \eqref{eq:learned_residual} telescope, and the remaining sum equals $J(\pi)$ by \eqref{eq:performance_cost}.
\end{proof}

\subsection{Proof of Comparison with Conventional Filtering}
\label{app:proof_improvement}
\begin{proof}
Applying Lemma~\ref{lem:residual_identity} to the conventional filter gives $J(\pi_{\mathrm{CBF}})=R_\theta(\pi_{\mathrm{CBF}})-W_\theta(x_N^{\mathrm{CBF}})+W_\theta(x_0)$. Subtracting this identity from the pathwise certificate \eqref{eq:pathwise_certificate} for the proposed filter, and using the common initial condition, yields \eqref{eq:performance_improvement_bound}. The condition \eqref{eq:residual_improvement_condition} makes the right-hand side strictly negative.
\end{proof}

\subsection{Supplementary Remarks}
\label{app:nonmyopic_remarks}

\begin{remark}[Computational structure]
\label{rem:computational}
The objective of \eqref{eq:nonmyopic_safety_filter} is quadratic, whereas the exact discrete-time barrier and learned constraints may be nonlinear in $u_k$. When $h$ is affine in the successor state, the barrier constraint is affine in $u_k$. The learned constraint is generally nonconvex, but can be approximated by linearizing $W_\theta$ about the nominal successor $\hat x_{k+1}=f(x_k)+g_u(x_k)u_{\mathrm{nom},k}$ when this successor lies in the differentiability domain of $W_\theta$,
\begin{equation}
\label{eq:linearized_learned_constraint}
W_\theta(x_{k+1})
\approx
W_\theta(\hat x_{k+1})
+
\nabla W_\theta(\hat x_{k+1})^\top
g_u(x_k)
(u_k-u_{\mathrm{nom},k}).
\end{equation}
Since the stage cost \eqref{eq:intervention_cost} remains convex quadratic, the resulting learned constraint is convex quadratic in $(u_k,\delta_k)$. With affine barrier constraints and a polyhedral input set $U$, the approximated problem is therefore a convex quadratically constrained quadratic program. Alternatively, the original problem may be solved as a nonlinear program warm-started at $u_k^{\mathrm{CBF}}$. In either case, the barrier constraint is imposed without approximation, so the safety guarantee remains independent of the treatment of the learned constraint. Performance certificates must, however, use the residual and relaxation evaluated for the original learned constraint.
\end{remark}

\begin{remark}[Recovery of the conventional filter]
\label{rem:recovery}
If $r_\theta(x_k,u_k^{\mathrm{CBF}})\le\eta$, the pair $(u_k^{\mathrm{CBF}},0)$ minimizes \eqref{eq:nonmyopic_safety_filter}. The CBF input already minimizes the intervention term, and its slack penalty is zero. If the CBF minimizer is unique, the proposed filter returns the same input. On a compact admissible input set with continuous learned residual, the NMF input also converges to the unique CBF minimizer as $\lambda\to0^+$.
\end{remark}

\begin{remark}[Relation to predictive safety filters]
\label{rem:relation_predictive}
Predictive CBFs use predictive information to characterize or approximate safety-related quantities~\cite{Wabersich2023,Didier2026}. Here, $W_\theta$ estimates cumulative intervention cost. The constructions can be combined by using a predictive barrier as $h$ in \eqref{eq:nonmyopic_safety_filter}.
\end{remark}

\begin{remark}[Relation to CLF-CBF quadratic programs]
\label{rem:clf_relation}
Like CLF-CBF quadratic programs~\cite{ames2014}, \eqref{eq:nonmyopic_safety_filter} combines a hard barrier constraint with a relaxed inequality. In this case, the relaxed Bellman residual is a dissipation inequality with storage $W_\theta$ and supply $-\ell(x_k,u_k)+\eta+\delta$. It guides cumulative intervention effort. Summing that inequality yields Lemma~\ref{lem:certificate}.
\end{remark}

\begin{remark}[Online certificate]
\label{rem:certificate_online}
The bound in \eqref{eq:pathwise_certificate} uses the learned endpoint values, the tolerance, and the realized slack. For measured successors under model mismatch or disturbances, the same telescoping argument applies if the residual inequality is evaluated and satisfied at those successors. Using only predicted successors requires accounting for the resulting difference in the learned value.
\end{remark}

\begin{remark}[Interpretation of the performance bound]
\label{rem:tightness}
Summing \eqref{eq:bellman_inequality} gives $J(\pi)\ge V^\star(x_0)-V^\star(x_N^\pi)$ on every barrier-admissible trajectory. For the proposed filter, \eqref{eq:approx_nonmyopic_bound} bounds the gap above this trajectory-dependent lower bound by $2\varepsilon_W+N\eta+\sum_{k=0}^{N-1}\delta_k^\star$.
\end{remark}

\subsection{Comparison with One-Step Learned-Value Lookahead}
\label{app:lookahead_comparison}
\begin{remark}[Non-accumulation of approximation error]
\label{rem:non_accumulation_extended}
The approximation error enters \eqref{eq:approx_nonmyopic_bound} through the endpoint term $2\varepsilon_W$, rather than through a term proportional to the horizon, because the learned constraint is summed before applying Assumption~\ref{ass:uniform_approximation}. For comparison, consider the one-step lookahead policy
\begin{equation}
\label{eq:lookahead_policy}
u_k^{\mathrm{LA}}
\in
\arg\min_{u\in K_{\mathcal C}(x_k)}
\left[
\ell(x_k,u)+W_\theta\!\left(f(x_k)+g_u(x_k)u\right)
\right].
\end{equation}
Let $u_k^{\mathrm{opt}}$ attain equality in \eqref{eq:bellman_inequality}. If the successors under $u_k^{\mathrm{LA}}$ and $u_k^{\mathrm{opt}}$ both belong to $\mathcal X_r$, optimality of $u_k^{\mathrm{LA}}$ and \eqref{eq:uniform_approximation} imply
\begin{equation}
\begin{aligned}
&\ell(x_k,u_k^{\mathrm{LA}})
+
V^\star(x_{k+1}^{\mathrm{LA}})\\
&\quad\le
\ell(x_k,u_k^{\mathrm{opt}})
+
V^\star(x_{k+1}^{\mathrm{opt}})
+
2\varepsilon_W\\
&\quad=
V^\star(x_k)+2\varepsilon_W.
\end{aligned}
\end{equation}
Summing along the lookahead trajectory gives $J(\pi_{\mathrm{LA}})\le V^\star(x_0)-V^\star(x_N^{\mathrm{LA}})+2N\varepsilon_W$, and hence $J(\pi_{\mathrm{LA}})\le J_N^\star(x_0)+2N\varepsilon_W$ under Assumption~\ref{ass:terminal_inactivity}. In contrast, the proposed filter yields a bound involving only endpoint approximation errors, but additionally includes the accumulated relaxation $\sum_k\delta_k^\star$. Since this relaxation is not bounded a priori, the comparison does not establish uniform superiority of either policy.
\end{remark}

\section{Additional Numerical Results and Implementation Details}
\label{app:results}

\subsection{Scenario and Exact Obstacle Constraints}
\label{app:results_setup}

The two spherical keep-out regions have centers $o_1=(0,1.0,1.2)^\top\,\mathrm m$ and $o_2=(3.5,-1.0,2.5)^\top\,\mathrm m$ and inflated radii $r_1=2.0\,\mathrm m$ and $r_2=1.8\,\mathrm m$. The inflation accounts for a vehicle radius of $0.20\,\mathrm m$. The command box $U=([-3.2,3.2]^2\times[-3,5])\,\mathrm{m/s^2}$ lies within the thrust and tilt limits of $16\,\mathrm N$ and $35^\circ$ for a $1\,\mathrm{kg}$ vehicle under ideal thrust-vector tracking. The saturated PD law in Section~\ref{sec:results} tracks the goal $p_{\mathrm g}=(7,0,1.7)^\top\,\mathrm m$. The $100$ test initial conditions, which are not used for training or parameter selection, are sampled independently over $p_x\in[-7.35,-6.65]$, $p_y\in[-1.4,1.4]$, $p_z\in[0.9,2.5]$ in meters and $v_x\in[0,3.3]$, $v_y,v_z\in[-0.4,0.4]$ in meters per second. A run reaches the goal if it completes all $N$ steps safely and ends within $0.50\,\mathrm m$ of $p_{\mathrm g}$ at a speed of at most $0.30\,\mathrm{m/s}$.

Let $d_i(p)=\|p-o_i\|_2-r_i$ be the clearance to the inflated obstacle. Define $n_i(p)=(p-o_i)/\|p-o_i\|_2$ and $h_i(x)=d_i(p)+\tau n_i(p)^\top v$ with $\tau=0.60\,\mathrm s$. Write $\bar p=p+\Delta t\,v$ and $\bar v=v-\Delta t\,c_d\|v\|_2v$. The exact sampled inequality $h_i(x_{k+1})\ge(1-\gamma)h_i(x_k)$, with $\gamma=0.20$, is implemented at $x_k=x$ as
\begin{equation}
\label{eq:sim_barrier}
-\tau\Delta t\,n_i(\bar p)^\top u_k
\le d_i(\bar p)+\tau n_i(\bar p)^\top\bar v
 -(1-\gamma)h_i(x),\qquad i=1,2.
\end{equation}
The rows are affine in $u_k$ because the Euler successor position is independent of the current command. In exact arithmetic, if $d_i(p)\ge0$ and $h_i(x)\ge0$, then for $s\in[0,\Delta t]$ convexity gives $d_i(p+sv)\ge d_i(p)+s\,n_i(p)^\top v\ge(1-s/\tau)d_i(p)\ge0$. The sampled barrier inequality then gives $h_i(x_{k+1})\ge0$. The simulator additionally checks sampled clearance and the minimum clearance along each straight segment, with a tolerance of $10^{-7}$. This argument concerns the discretized model and its position segments.

\subsection{Training Data and Offline Optimal Solution}
\label{app:results_training}

The training data use $533$ starts from the test distribution and $267$ from a wider region. The $100$ test initial conditions and $40$ validation starts are separate from the training starts. Labels are generated at the initial states and at states visited by the conventional CBF filter and by optimized continuations. Near an obstacle, the dataset also includes pairs of feasible successor states reached by different first commands. The members of each pair share position and differ in velocity. States inside the terminal ellipsoid $\mathcal X_{\mathrm f}=\{x:(x-x_{\mathrm g})^\top P(x-x_{\mathrm g})\le\rho\}$, with $x_{\mathrm g}=(p_{\mathrm g},0)$ and $\rho\simeq0.82525$, receive zero labels. For a label state $x$ outside $\mathcal X_{\mathrm f}$, the offline solver uses the terminal-constrained problem
\begin{equation}
\label{eq:quadrotor_label_problem}
\begin{aligned}
\min_{u_0,\ldots,u_{89}}\quad&
\sum_{k=0}^{89}\ell(x_k,u_k)\\
\textnormal{s.t.}\quad&
x_0=x,\qquad x_{90}\in\mathcal X_{\mathrm f},\\
&x_{k+1}\text{ satisfies \eqref{eq:quadrotor_dynamics}},\\
&u_k\in U,\qquad \eqref{eq:sim_barrier},\qquad k=0,\ldots,89.
\end{aligned}
\end{equation}
The label is the cost of the lowest-cost feasible candidate found. The code checks sufficient conditions for $\mathcal X_{\mathrm f}$ to be invariant under $\kappa$ and for $\kappa$ to satisfy the input and barrier constraints there. Thus, from $x_{90}\in\mathcal X_{\mathrm f}$, applying $u_k=\kappa(x_k)$ for all $k\ge90$ gives a feasible continuation with zero further intervention cost.

The offline label solver generates feasible initializations by filtering the nominal law and waypoint-tracking laws. It refines one initialization per solve using SciPy SLSQP. Three networks are trained on the same data with different random initializations (seeds $11$, $22$, and $33$). Each uses standardized goal-relative inputs, two hidden layers of $64$ $\tanh$ units, and a softplus output. Training minimizes squared error on scaled square-root targets using Adam with learning rate $3\cdot10^{-3}$, batch size $128$, and $800$ epochs. The rescaled prediction is multiplied by a continuously differentiable gate and squared to obtain $W_\theta$. The gate is zero inside $\mathcal X_{\mathrm f}$. The dataset contains $2{,}908$ zero targets. The seed-$11$ network is used to choose $(\lambda,\eta)$ on the validation set, with candidates ranked by safe completions, goal arrivals, and then mean cost among goal-reaching runs.

For each test start, the offline optimal solution is the lowest-cost feasible trajectory among the cached plan and the safe controller rollouts that complete all $N=90$ steps, including auxiliary runs. Its cost $\widehat J_N$ is an upper bound on the finite-horizon optimum $J_N^\star(x_0)$ for the implemented constraints. A separate terminal-constrained comparison retains candidates that end in $\mathcal X_{\mathrm f}$. The label error is measured against the cached terminal-constrained plans before including the controller rollouts. Replaying the $100$ cached test plans gives mean absolute errors of $1.62$, $1.69$, and $1.88$ for seeds $11$, $22$, and $33$, respectively. For seed $11$, the $95$th percentile and maximum absolute errors are $5.81$ and $20.55$. These are empirical errors relative to numerical labels for the $100$ test initial conditions, not bounds on the error relative to $V^\star$. The terminal construction gives a sufficient condition for Assumption~\ref{ass:terminal_inactivity} if a finite-horizon optimizer reaches $\mathcal X_{\mathrm f}$.

\subsection{Closed-Loop Intervention and Safety Histories}
\label{app:results_trajectories}

\begin{figure}[t]
\centering
\includegraphics[width=\columnwidth]{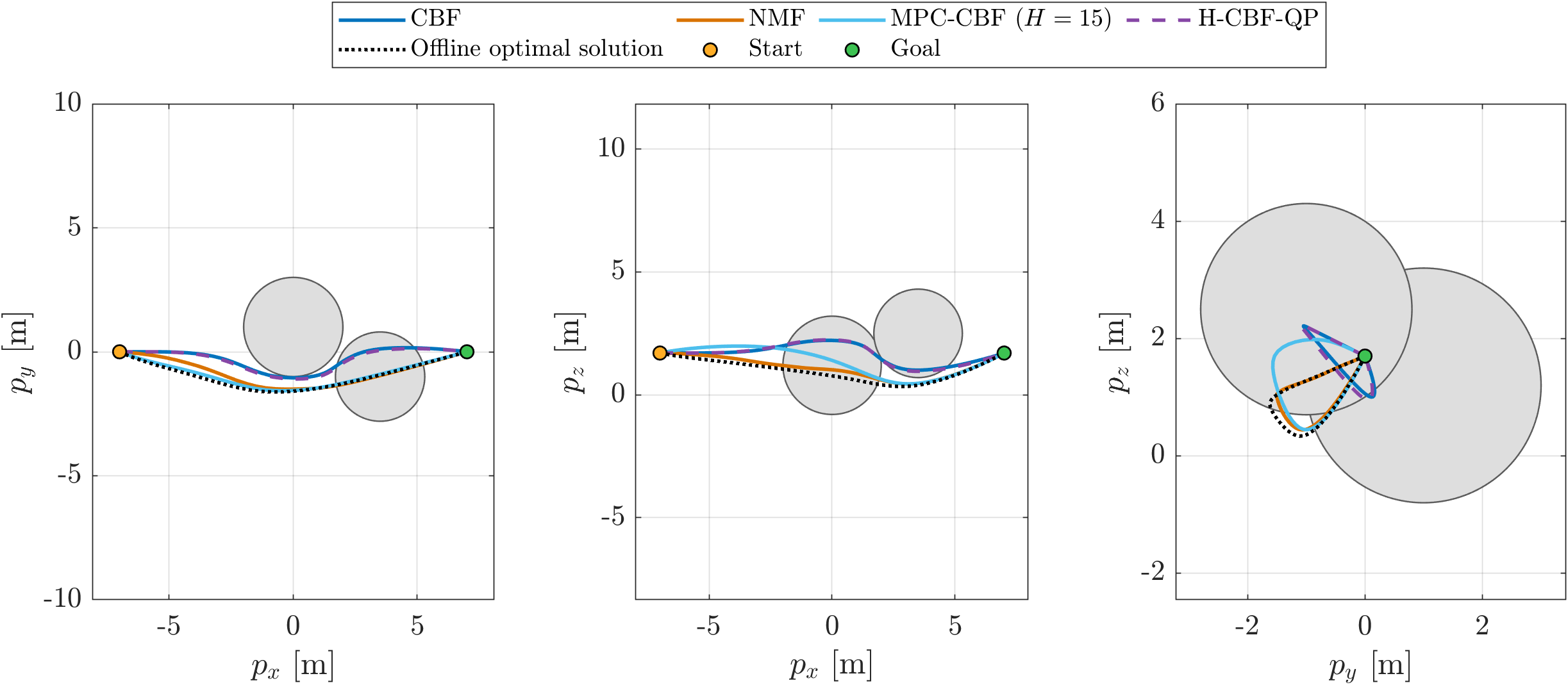}
\caption{Orthogonal projections of the trajectories in Fig.~\ref{fig:trajectory}.}
\label{fig:quadrotor_planar}
\end{figure}

\begin{figure}[t]
\centering
\includegraphics[width=0.6\columnwidth]{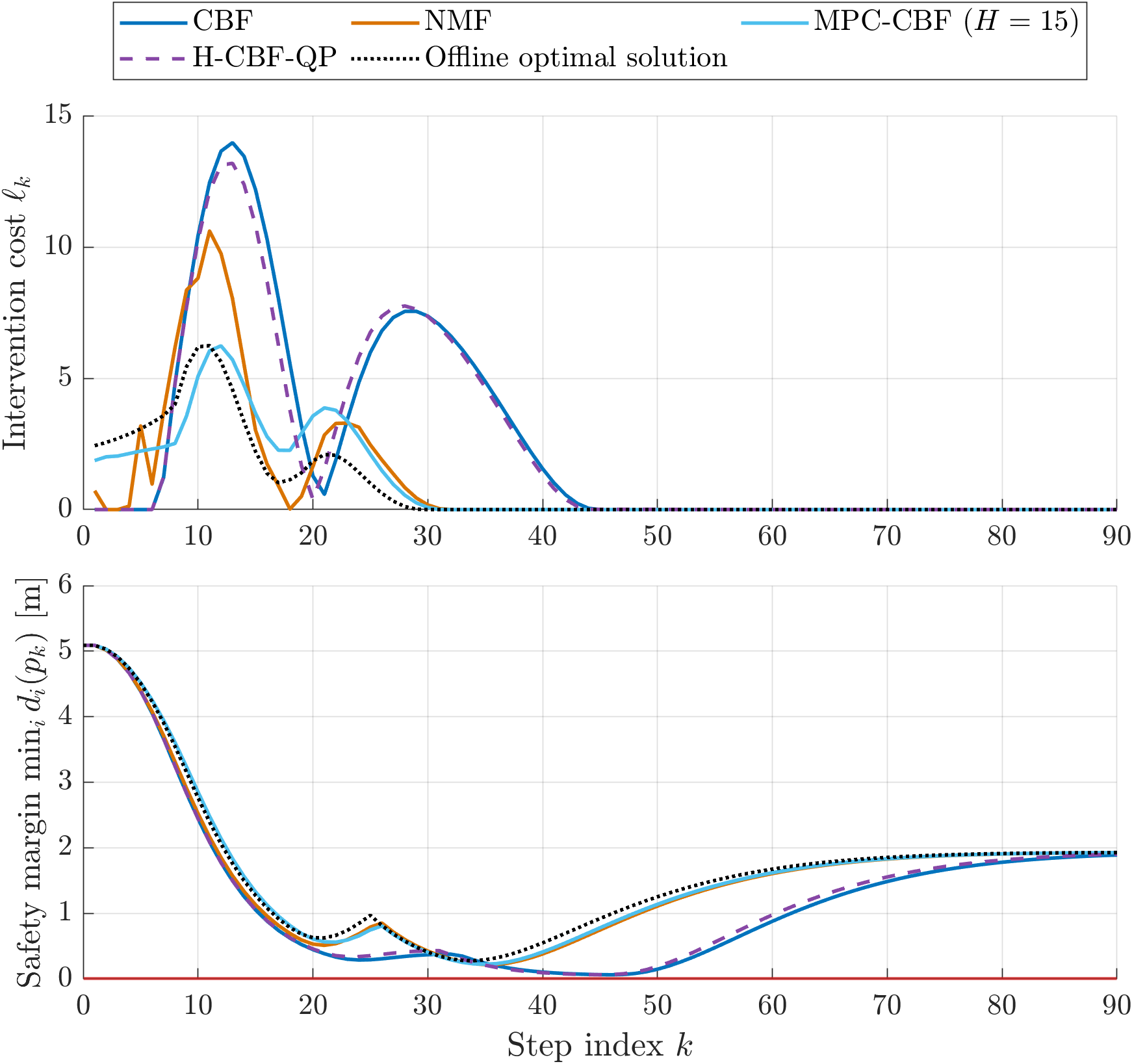}
\caption{Stage intervention cost and sampled obstacle clearance.}
\label{fig:cost_margin}
\end{figure}

\begin{figure}[t]
\centering
\includegraphics[width=0.6\columnwidth]{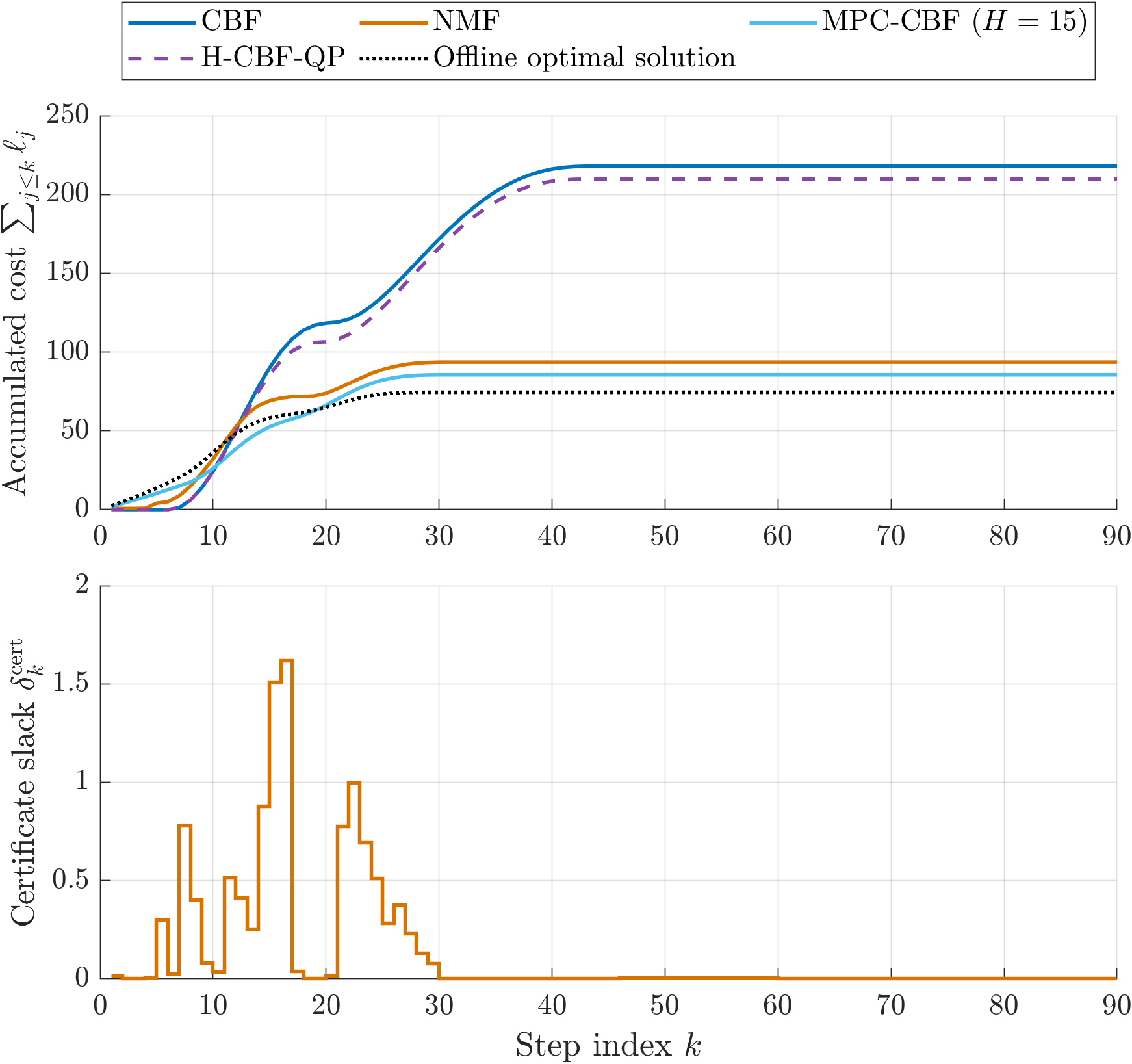}
\caption{Cumulative intervention cost and NMF slack.}
\label{fig:accum_slack}
\end{figure}

The projections in Fig.~\ref{fig:quadrotor_planar} show that the NMF path has a lateral deviation similar to that of the offline optimal solution. In Fig.~\ref{fig:cost_margin}, the peak stage costs are approximately $14.0$ for CBF, $13.2$ for H-CBF-QP, $10.6$ for NMF, and $6.2$ for MPC-CBF. The minimum sampled clearances in the same order are $0.062$, $0.062$, $0.217$, and $0.222\,\mathrm m$. The corresponding minimum clearances over the straight segments between samples are $0.060$, $0.060$, $0.215$, and $0.221\,\mathrm m$, computed from the saved positions. Figure~\ref{fig:accum_slack} shows that the largest NMF slack, about $1.62$, occurs at plotted step $16$. Most slack accrues during the first $30$ steps and quantifies the positive excess of the learned residual above $\eta$.

\subsection{Controller Implementation and Execution Time}
\label{app:results_timing}

The conventional filter projects $\kappa(x_k)$ onto the common input set defined by $U$ and \eqref{eq:sim_barrier}. H-CBF-QP projects the Hamiltonian minimizer $-\tfrac12\nabla_v V_{\mathrm H}(x_k)$ onto the same set. The approximate value $V_{\mathrm H}$ is fitted offline for an unconstrained quadratic tracking and effort cost by least-squares policy iteration. Both projections use MATLAB's \texttt{quadprog} when their targets are inadmissible.

NMF eliminates the scalar slack analytically and minimizes $\tfrac12\ell(x_k,u)+\tfrac{\lambda}{2}[r_\theta(x_k,u)-\eta]_+^2$ subject to the common input constraints. It returns the nominal input directly if it satisfies both the hard constraints and the residual threshold. Otherwise it uses MATLAB's \texttt{fmincon} with SQP and retains the conventional CBF input as a feasible incumbent. The implementation evaluates the original nonlinear residual and supplies its analytic gradient. MPC-CBF uses the same dynamics, input bounds, and obstacle-wise barrier inequalities along its $15$-step prediction. It warm-starts SQP from the shifted previous plan when available and uses the checked CBF input if no acceptable prediction is found.

\begin{figure}[t]
\centering
\includegraphics[width=\columnwidth]{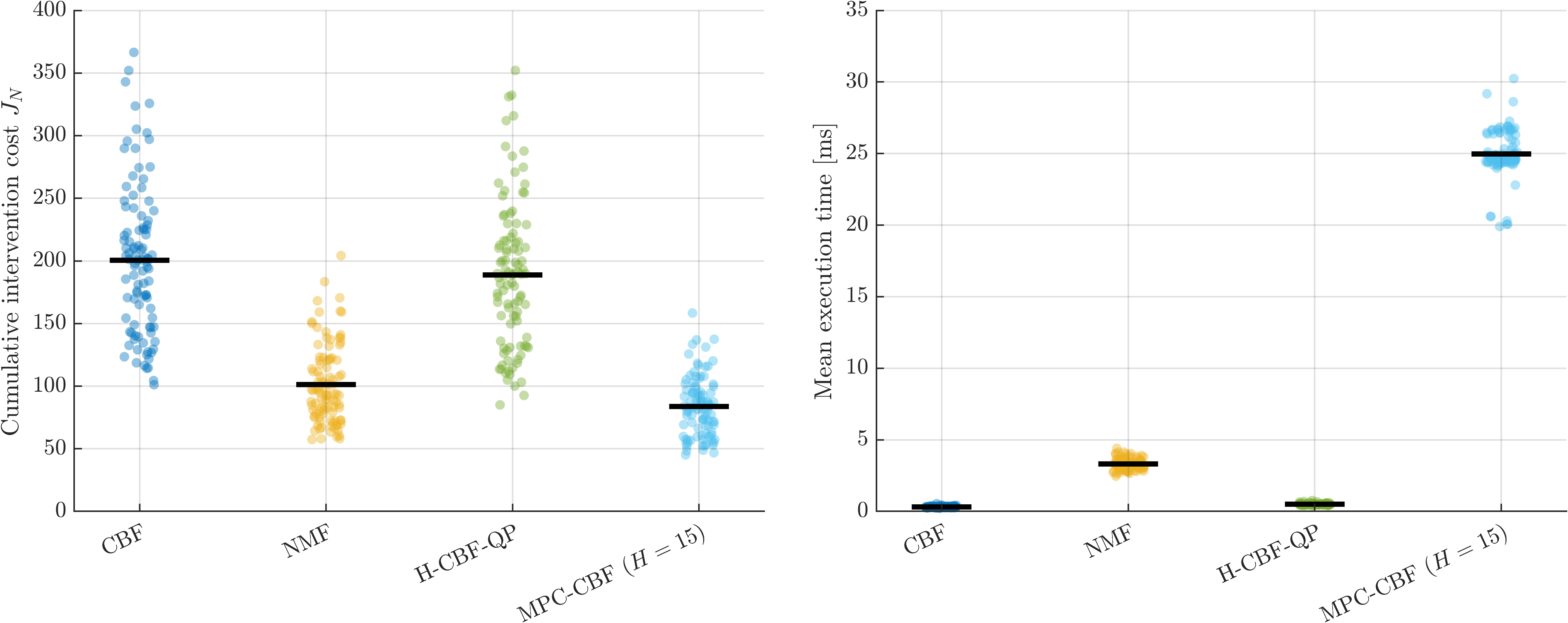}
\caption{Cumulative intervention costs and mean computation times over $100$ test cases.}
\label{fig:exec_time}
\end{figure}

Figure~\ref{fig:exec_time} shows the cumulative intervention costs and mean online computation times per step for the $100$ test initial conditions using the first network (seed $11$). Each marker represents one test case and the horizontal lines indicate the means. Mean intervention costs are $200.50$ for CBF, $188.78$ for H-CBF-QP, $101.26$ for NMF, and $83.76$ for MPC-CBF. NMF has lower cumulative intervention cost than CBF and H-CBF-QP in every test case. All four controllers complete the $90$-step runs safely and reach the goal. The NMF reductions relative to CBF are $50.6\,\%$ and $49.2\,\%$ for the other two independently trained networks.

Mean and observed maximum online times are $(0.31,6.97)$ for CBF, $(0.50,10.78)$ for H-CBF-QP, $(3.31,37.69)$ for NMF, and $(24.97,81.32)\,\mathrm{ms}$ for MPC-CBF. The measured times include direct returns and fallback calls when taken. The online decision dimension is $3$ for the slack-eliminated NMF problem and $45$ for the MPC-CBF input sequence. The maxima are specific to the recorded runs.

\subsection{Evaluation of the A Posteriori Improvement Condition}
\label{app:results_certificate}

For each common start with two complete $N$-step trajectories, evaluate the learned residual with the network used by NMF and reconstruct the realized slack as $\delta_k^\star=[r_\theta(x_k^{\mathrm{NMF}},u_k^{\mathrm{NMF}})-\eta]_+$. With $\Delta_W=W_\theta(x_N^{\mathrm{CBF}})-W_\theta(x_N^{\mathrm{NMF}})$, the sufficient condition of Theorem~\ref{thm:performance_improvement} is
\begin{equation}
\label{eq:app_improvement_check}
R_\theta(\pi_{\mathrm{CBF}})>\Delta_W+N\eta+\sum_{k=0}^{N-1}\delta_k^\star.
\end{equation}
By telescoping the CBF residuals, this check is equivalent to comparing $J(\pi_{\mathrm{CBF}})$ with the NMF certificate
\[
B_\theta=W_\theta(x_0)-W_\theta(x_N^{\mathrm{NMF}})+N\eta+\sum_{k=0}^{N-1}\delta_k^\star .
\]
The benchmark stores $B_\theta$ for each completed NMF rollout. Cases with $J(\pi_{\mathrm{CBF}})>B_\theta$ certify strict improvement, since $J(\pi_{\mathrm{NMF}})\le B_\theta$. Cases with $J(\pi_{\mathrm{NMF}})<J(\pi_{\mathrm{CBF}})\le B_\theta$ are observed improvements for which this sufficient test is inconclusive.
For the run from rest, the saved network gives $W_\theta(x_0)\simeq89.72$, and the reconstructed penultimate state has $W_\theta(x_{N-1}^{\mathrm{NMF}})=0$. The last plotted slack is zero to plotting precision. Since $\ell$ and $W_\theta$ are nonnegative and $\eta=0$, this gives $W_\theta(x_N^{\mathrm{NMF}})\simeq0$. The plotted slack sums to approximately $10.95$. With $\eta=0$, the resulting certificate is $B_\theta\simeq100.67$, compared with $J(\pi_{\mathrm{NMF}})\simeq93.60$ and $J(\pi_{\mathrm{CBF}})\simeq217.98$. The sufficient condition is therefore satisfied for this start.

We also evaluate the certificate for all $100$ test initial conditions and the three independently trained networks. For seeds $11$, $22$, and $33$, the mean realized NMF costs are $101.26$, $99.14$, and $101.92$, respectively, while the corresponding mean certificate values are $101.44$, $99.55$, and $102.30$. The sufficient condition in \eqref{eq:app_improvement_check} certifies strict improvement over CBF in $100$, $99$, and $100$ cases, respectively. The remaining case also has a lower realized NMF cost than CBF, although the sufficient condition is inconclusive.

%% file: references.bib
@inproceedings{Agrawal2017,
  author    = {Agrawal, Ayush and Sreenath, Koushil},
  title     = {Discrete Control Barrier Functions for Safety-Critical
               Control of Discrete Systems with Application to Bipedal
               Robot Navigation},
  booktitle = {Proceedings of Robotics: Science and Systems}, 
  doi={10.15607/RSS.2017.XIII.073},
  volume    = {13},
  address   = {Cambridge, MA, USA},
  year      = {2017}
}

@misc{ShahrakiLessard2026,
  title         = {Constrained Feedback Control of Nonlinear Systems
                   via Approximate {HJB} and Control Barrier Functions},
  author        = {Alipour Shahraki, Milad and Lessard, Laurent},
  year          = {2026},
  eprint        = {2603.16114},
  archivePrefix = {arXiv},
  primaryClass  = {eess.SY},
  doi           = {10.48550/arXiv.2603.16114},
  url           = {https://arxiv.org/abs/2603.16114v2},
  note          = {Version 2}
}

@inproceedings{ames2014,
  author    = {Ames, Aaron D. and Grizzle, Jessy W. and Tabuada, Paulo},
  title     = {Control Barrier Function Based Quadratic Programs with
               Application to Adaptive Cruise Control},
  booktitle = {2014 53rd IEEE Conference on Decision and Control (CDC)},
  doi={10.1109/CDC.2014.7040372},
  year      = {2014}
}

@INPROCEEDINGS{Ames_2019,
author={Ames, Aaron D. and Coogan, Samuel and Egerstedt, Magnus and Notomista, Gennaro and Sreenath, Koushil and Tabuada, Paulo},
booktitle={2019 18th European Control Conference (ECC)}, 
title={Control Barrier Functions: Theory and Applications}, 
year={2019},
volume={},
number={},
pages={3420-3431},
doi={10.23919/ECC.2019.8796030}
}

@article{autenrieb2025_JGCD,
  author  = {Autenrieb, Johannes},
  title   = {Quadratic Programming Approach to Flight Envelope Protection
             Using Control Barrier Functions},
  journal = {Journal of Guidance, Control, and Dynamics},
  year    = {2025},
  month   = sep,
  doi     = {10.2514/1.G009203}
}

@inproceedings{autenrieb2025_ACC,
  author    = {Autenrieb, Johannes and Shin, Hyo-Sang},
  title     = {Sensor-Based Safety-Critical Control Using an Incremental
               Control Barrier Function Formulation via Reduced-Order
               Approximate Models},
  booktitle = {2025 American Control Conference (ACC)},
  doi={10.23919/ACC63710.2025.11107913},
  month     = {jul},
  year      = {2025}
}

@article{Blanchini_1999,
title = {Set invariance in control},
journal = {Automatica},
volume = {35},
number = {11},
pages = {1747-1767},
year = {1999},
issn = {0005-1098},
doi = {https://doi.org/10.1016/S0005-1098(99)00113-2    },
author = {F. Blanchini}
}

@ARTICLE{Didier2026,
  author={Didier, Alexandre and Zeilinger, Melanie N.},
  journal={IEEE Transactions on Automatic Control}, 
  title={Approximate Predictive Control Barrier Function for Discrete-Time Systems}, 
  year={2026},
  volume={71},
  number={7},
  doi={10.1109/TAC.2026.3662563}
  }

@article{Didier2024,
  author  = {Didier, Alexandre and Zanelli, Andrea and Wabersich, Kim P.
             and Zeilinger, Melanie N.},
  title   = {Predictive Stability Filters for Nonlinear Dynamical Systems
             Affected by Disturbances},
  journal = {IFAC-PapersOnLine},
  doi = {10.1016/j.ifacol.2024.09.031},
  volume  = {58},
  number  = {18},
  year    = {2024}
}

@article{GrosZanonBemporad2020,
  author  = {Gros, Sebastien and Zanon, Mario and Bemporad, Alberto},
  title   = {Safe Reinforcement Learning via Projection on a Safe Set:
             How to Achieve Optimality?},
  journal = {IFAC-PapersOnLine},
  year    = {2020},
  volume  = {53},
  number  = {2},
  pages   = {8076--8081},
  url     = {https://arxiv.org/abs/2004.00915}
}

@inproceedings{Hafner2026,
  author    = {Hafner, S. and Autenrieb, Johannes and Steinert, A. and
               Holzapfel, Florian},
  title     = {Control Barrier Functions on a Pseudo Control Level for
               Safe Flight Control},
  booktitle = {Proceedings of the 2026 CEAS EuroGNC Conference},
  address   = {Madrid, Spain},
  month     = may,
  year      = {2026},
  doi = {10.82124/CEAS-GNC-2026-010}
}

@article{LinXu2026,
  author  = {Lin, Ziyan and Xu, Liang},
  title   = {Curvature-Guided Safety Filters: State-Dependent
             {Hessian}-Weighted Projection With Provable Performance Bounds},
  journal = {IEEE Control Systems Letters},
  year    = {2026},
  volume  = {12},
  pages   = {139--144},
  doi     = {10.1109/LCSYS.2026.3672437}
}

@ARTICLE{Lindemann2024,
  author={Lindemann, Lars and Robey, Alexander and Jiang, Lejun and Das, Satyajeet and Tu, Stephen and Matni, Nikolai},
  journal={IEEE Open Journal of Control Systems}, 
  title={Learning Robust Output Control Barrier Functions From Safe Expert Demonstrations}, 
  year={2024},
  volume={3},
  number={},
  doi={10.1109/OJCSYS.2024.3385348},
  pages={158-172}
  }

@article{Shi2024,
  author  = {Shi, Kaixiang and Hu, Guoqiang},
  title   = {Safety-Guaranteed and Task-Consistent Human-Robot Interaction
             Using High-Order Time-Varying Control Barrier Functions and
             Quadratic Programs},
  journal = {IEEE Robotics and Automation Letters},
  volume  = {9},
  number  = {1},
  pages   = {547--554},
  doi={10.1109/LRA.2023.3333166},
  year    = {2024}
}

@ARTICLE{Wabersich2023,
  author={Wabersich, Kim P. and Zeilinger, Melanie N.},
  journal={IEEE Transactions on Automatic Control}, 
  title={Predictive Control Barrier Functions: Enhanced Safety Mechanisms for Learning-Based Control}, 
  year={2023},
  volume={68},
  doi={10.1109/TAC.2022.3175628},
  number={5}
  }

@article{wabersich2021predictive,
title = {A predictive safety filter for learning-based control of constrained nonlinear dynamical systems},
journal = {Automatica},
volume = {129},
year = {2021},
doi = {https://doi.org/10.1016/j.automatica.2021.109597},
author = {Kim Peter Wabersich and Melanie N. Zeilinger}
}

@ARTICLE{Xiong2023,
  author={Xiong, Yuhan and Zhai, Di-Hua and Tavakoli, Mahdi and Xia, Yuanqing},
  journal={IEEE Transactions on Cybernetics}, 
  title={Discrete-Time Control Barrier Function: High-Order Case and Adaptive Case}, 
  year={2023},
  volume={53},
  number={5},
  pages={3231-3239},
  doi={10.1109/TCYB.2022.3170607}}

@INPROCEEDINGS{Zeng2021,
  author={Zeng, Jun and Li, Zhongyu and Sreenath, Koushil},
  booktitle={2021 60th IEEE Conference on Decision and Control (CDC)}, 
  title={Enhancing Feasibility and Safety of Nonlinear Model Predictive Control with Discrete-Time Control Barrier Functions}, 
  year={2021},
  volume={},
  number={},
  pages={6137-6144},
  doi={10.1109/CDC45484.2021.9683174}}
